\documentclass[12pt,preprint]{aastex}
\usepackage{apjfonts}  

\usepackage{epsfig,graphicx,latexsym,amsmath,amssymb}
\usepackage{natbib}
\usepackage{hyperref}
\usepackage{mathrsfs}
\usepackage{booktabs}

\bibpunct[,]{(}{)}{;}{a}{}{,}

\begin{document}

\author{Pau Amaro-Seoane\altaffilmark{1}\thanks{e-mail: Pau.Amaro-Seoane@aei.mpg.de} \&
Luc\'{i}a Santamar\'{i}a\altaffilmark{2}\thanks{e-mail: Lucia.Santamaria@aei.mpg.de}}
\altaffiltext{1}
{(PAS) Max Planck Institut f\"ur Gravitationsphysik
(Albert-Einstein-Institut), D-14476 Potsdam, Germany and Institut de Ci{\`e}ncies de l'Espai, IEEC/CSIC, Campus UAB,
Torre C-5, parells,
$2^{\rm na}$ planta, ES-08193, Bellaterra, Barcelona, Spain}
\altaffiltext{2}
{(LS) Max Planck Institut f\"ur Gravitationsphysik
(Albert-Einstein-Institut), D-14476 Potsdam, Germany}

\date{\today}

\title{Detection of IMBHs with ground-based
  gravitational wave observatories:\\ 
       A biography of a binary of black holes, from birth to death}

\begin{abstract}
Even though the existence of intermediate-mass black holes (IMBHs, black holes
with masses  ranging between $10^{2-4}\,M_{\odot}$) has not yet been corroborated
observationally, these objects are of high interest for astrophysics. Our
understanding of the formation and evolution of supermassive black
holes (SMBHs), 
as well as galaxy evolution modeling and cosmography would dramatically change
if an IMBH were to be observed. From a point of view of traditional photon-based
astronomy, {which relies on the monitoring of innermost stellar
  kinematics}, the {\em direct} detection of an IMBH seems to be
rather far in the future. 
However, the prospect of the detection and
characterization of an IMBH has good chances in lower-frequency
gravitational-wave (GW) astrophysics using ground-based detectors such as
LIGO, Virgo and the future Einstein Telescope (ET).
We present an analysis of the signal of a system of a binary of IMBHs
(BBH from now onwards) based on a waveform model obtained with
numerical relativity simulations coupled with post-Newtonian calculations at
the highest available order.
IMBH binaries with total masses between $200-20000\,M_\odot$ 
would produce significant signal-to-noise ratios (SNRs) in advanced LIGO
and Virgo and the ET. 
We have computed the expected event rate of IMBH binary
coalescences for different configurations of the binary, finding
interesting values that depend on the spin of the IMBHs.
The prospects for IMBH detection and characterization with ground-based GW
observatories would not only provide us with a robust test of general
relativity, but would also corroborate the existence of these
systems. 
Such detections should allow astrophysicists to probe the stellar
environments of IMBHs and their formation processes.

\end{abstract}

\maketitle

\section{Motivation}
\label{sec.motivation}

By following the stellar dynamics at the center of our Galaxy, we have now the
most well-established evidence for the existence of a SMBH.  The close
examination of the Keplerian orbits of the so-called S-stars (also called
S0-stars, where the letter ``S'' stands simply for source) has revealed the
nature of the central dark object located at the Galactic Center. By following
S2 (S02), the mass of SgrA$^*$ was estimated to be about $3.7\times
10^6\,M_{\odot}$ 
within a volume with radius no larger than 6.25 light-hours
\citep{SchoedelEtAl03,GhezEtAl03b}. More recent data based on 16 years of
observations set the mass of the central SMBH to $\sim 4
\times 10^{6} \, M_{\odot}$
\citep{EisenhauerEtAl05,GhezEtAl05,GhezEtAl08,GillessenEtAl09}.

Massive black holes in a lower range of masses may exist in smaller stellar
systems such as globular clusters. These are called intermediate-mass black
holes (IMBHs) because their masses range between $M\sim 10^{2}$ and $M\sim 10^{4}\,M_\odot$, 
assuming that they follow the observed correlations between SMBHs and their
host stellar environments. Nevertheless, 
the existence of IMBHs has never been confirmed, though
we have some evidence that could favor their existence \citep[see][ and references
therein]{MillerColbert04,Miller09}.   

If we wanted to apply the same technique to detect IMBHs in globular
clusters as we use for SMBHs in galactic centers, we would need ultra-precise
astronomy, since the sphere of influence of an IMBH is {$\sim$ few arc
seconds. For instance, for a $10^4\,M_{\odot}$ IMBH, the influence radius is of
$\sim 5''$ assuming a central velocity dispersion of $\sigma =
20\,{\rm km\,s}^{-1}$ and a distance of $\sim 5$ kpc.} The number of stars enclosed in
that volume is only a few. Currently, with adaptive optics, one can aspire
--optimistically-- to have a couple of measurements of velocities if the
target is about $\sim 5$ kpc away on a time scale of about 10 yrs. The measures
depend on a number of factors, such as the required availability of a bright
reference star in order to have a good astrometric reference system.  Also,
the sensitivity limits correspond to a K-band magnitude of $\sim$ 15, (B- MS
stars at 8 kpc, like e.g. S2 in our Galactic Center).

This means that, in order to detect an IMBH or, at least, a massive dark object
in a globular cluster center by following the stellar dynamics around it, one
has to have recourse to the Very Large Telescope interferometer and to one of the
next-generation instruments, the VSI or GRAVITY
\citep{GillessenEtAl06,EisenhauerEtAl08}. In this case we can hope to improve
the astrometric accuracy by a factor of $\sim 10$. Only in that scenario 
would we be in the position of following closely the kinematics around a potential
IMBH so as to determine its mass. 

GW astronomy could contribute to IMBH detection. 
In the past years, the field has reached a
milestone with the construction of an international network of GW
interferometers that have achieved or are close to achieving their
design sensitivity. 
Moreover, the first-generation ground-based detectors LIGO and Virgo will
undergo major technical upgrades in the next five years that will increase the
volume of the observable universe by a factor of a thousand 
\footnote{\scriptsize{http://www.ligo.caltech.edu/advLIGO/,
  http://wwwcascina.virgo.infn.it/advirgo/}}.  

The data that will be taken by the advanced interferometers are expected to
transform the field from GW detection to GW astrophysics. The availability of
accurate waveform models for the full BBH coalescence in order to construct
templates for match-filtering is crucial in the GW searches
for compact binaries.  The construction of this kind of templates has
recently been made possible thanks to the combination of post-Newtonian
calculations of the BBH inspiral and numerical relativity simulations of the
merger and ringdown. Two approaches to this problem are the
effective-one-body techniques~\citep{Buonanno:1998gg,Buonanno:2009qa}
and the phenomenological matching of PN 
and NR
results~\citep{Ajith:2007kx,Ajith:2009bn,Santamaria:SpinTemp}. In this
paper we use the results of the latter.  

The structure of this paper is as follows: In section~\ref{sec:astro} we expand
the astrophysical context to this problem and give a description of the
different efforts made to address the evolution of a BBH in a stellar cluster
from its birth to its final coalescence.  In section~\ref{sec:wfmodel} we
introduce the techniques used in the data analysis of the waveform
modeling of BBH coalescences, present our hybrid waveform model,
and use the model to compute and discuss expected signal-to-noise
ratios in present and future GW detectors.
The use of the new waveform model allows us to give an improved
estimate for the  number of events one can expect for several physical
configurations in 
Advanced LIGO and the Einstein Telescope, which we present in
section~\ref{sec:events}.
We conclude with a summary of 
our results and future prospects of our work in section~\ref{sec:concl}.

\section{Life of a massive binary}
\label{sec:astro}

The aim of this section is not to give a detailed explanation of the processes
of formation of IMBHs and BBHs, but a general introduction 
of the two different scenarios that play a role
in the formation of BBHs. 

\subsection{Birth}

Up to now, the IMBH formation process which has drawn the most
attention is that of 
a young cluster in which the most massive stars sink down to the center due to
mass segregation. There, a high-density stellar region builds and stars start
to physically collide. One of them gains more and more mass and forms a runaway
star whose mass is much larger than that of any other star in the
system, a very massive star (VMS).
Later, that runaway star may collapse and form an IMBH
\citep{PortegiesZwartMcMillan00,GurkanEtAl04,PortegiesZwartEtAl04,
FreitagEtAl06}. 

In particular, \cite{FreitagRB06,FreitagEtAl06} described in detail the
requirements from the point of view of the host cluster to form an IMBH in the
center of such a system. The cluster cannot have ``too hard''
binaries, the time to reach core-collapse must be shorter than 3~Myr
and the environmental velocity dispersion cannot be much larger than
$\sim 500\,{\rm yr}^{-1}$. Under these conditions the authors find
that the mass of the VMS formed is $\gg 100\,M_\odot$. 

The later evolution of the VMS is not well understood, nor are the necessary conditions 
that it not evolve into a super-massive star (SMS) \citep[see for
instance][ and the references in their work]{AS01,ASEtAl02,PauTesi04} in this
particular scenario, nor are the factors that could limit the mass of such
an object such that it not collapse into an IMBH. The process depends
on a number of factors and assumptions, such as e.g. the role of metallicity,
winds \citep[see e.g.][ though it is rather unclear how to extrapolate the
results they obtain, which are limited to stars with masses of maximum
$150\,M_{\odot}$ to the masses found in the runaway scenario, which are
typically at least one order of magnitude larger]{BelkusEtAl07} and the
collisions on to the runaway star from a certain mass upwards.  On the other
hand, \cite{SuzukiEtAl07} investigated the growth of a runaway
particle by coupling direct $N-$body simulations with smooth particle
hydrodynamics, (SPH) analyzed the evolution of the star and found that stellar
winds would not inhibit the formation of a very massive
star. More recently,
\cite{GlebbeekEtAl09} considered the effects of stellar evolution on the
runaway collision product by analyzing the succession of collisions from a
dynamical evolution. For their low-metallicity models, the final
remnant of the merger tree is expected to explode as a supernova, and in their
high-metallicity models the possibility of forming an IMBH is negligible,
finishing with a mass of 10--14 $M_{\odot}$ at the onset of carbon burning.
But as a matter of fact, these stars develop an extended envelope, so
that the probability of 
further collisions is higher. \cite{GlebbeekEtAl09} did not change the
masses in the dynamical simulation accordingly. In any case, self-consistent
direct-summation $N$-body simulations with evolution of the runaway process are
called in to investigate the final outcome.  Henceforth we will assume
that IMBHs do form; in that case, the formation of a binary of them
(BBH) in a cluster can be theoretically explained in two different
ways. 

{\em (i) The double-cluster channel: }

In this scenario, two clusters born in a cluster of clusters such as those
found in the Antenn{\ae} galaxy are gravitationally bound and doomed to
collide \citep[see][ for a detailed explanation of the process and their
references]{ASF06}. When this happens, the IMBHs sink down to the center of the
resulting merged stellar system due to dynamical friction. They form a BBH
whose semi-major axis continues to shrink due to slingshot ejections of stars
coming from the stellar system. In each of the processes, a star removes a
small fraction of the energy and angular momentum of the BBH, which becomes
harder. At later stages in the evolution of the BBH, GW radiation takes over
efficiently and the orbit starts to circularize, though one can expect these systems to
have a residual eccentricity when entering the LISA band \citep{ASF06,AS10a}.
This detector will typically be able to see systems of binaries of IMBHs out to
a few Gpc. For this channel and volume, the authors estimated an event rate of
$4-5~{\rm yr}^{-1}$. 

{\em (ii) The single-cluster channel: }

\cite{GFR06} added a fraction of
primordial binaries to the initial configuration in the scenario of formation
of a runaway star in a stellar cluster. In their simulations they find that not
one, but two very massive stars form in rich clusters with a binary fraction of
10\%. \cite{FregeauEtAl06} investigated the possibility of emission of GWs by
such a BBH and estimated that LISA and Advanced LIGO can detect tens of them
depending on the distribution of cluster masses and their densities. More
recently, \cite{Gair:2009gr} recalculated the rate of \cite{FregeauEtAl06} for
the case of the proposed Einstein Telescope could see and quoted a few to a few
thousand events of comparable-mass BBH mergers of the single-cluster channel.

\subsection{Growing up (shrinking down): The role of triaxiality on
  centrophilic orbits} 

In the case of the double-cluster channel, the
cluster, which is in rotation, results from the merger of the two initial
clusters and may develop a triaxiality sufficient to produce sufficient
centrophilic orbits. These ``boxy'' orbits, as seen by \cite{BerczikEtAl06},
are typical of systems that do not possess symmetry around any of their axes.
In contrast to loop orbits, a characteristic of spherically symmetric or
axisymmetric systems, ``boxy'' orbits bring stars arbitrarily close to the
center of the system, since it oscillates independently along the three
different axes.  Therefore, such stars, due to the fact of being potential
sling-shots, can feed the process of shrinkage of the BBH semi-major axes by
removing energy and angular momentum from it after a strong interaction. In
the strong triaxial systems of \cite{BerczikEtAl06}, the rotation caused in the
process of merger creates an unstable structure in the form of a bar. Within
the bar the angular momentum will not be conserved and thus the BBH loss-cone
is full due to stars on centrophilic orbits, independently of the number of
stars $N_{\star}$. On the other hand, for BBH systems in clusters, the role of
Brownian motion has an impact in the loss-cone \citep[see][]{AS10a}.
In the models of \cite{ASF06}, the initial conditions are a realistic parabolic
merger of two stellar clusters. The resulting merged cluster does not show the
strong axisymmetry of \cite{BerczikEtAl06}.  
In the simulations we address for the results of this work, the BBH (of
IMBHs) is not stalling, in spite of the
reduced number of centrophilic orbits due to the architecture of the stellar
system.

\begin{figure}
\resizebox{\hsize}{!}{\includegraphics[clip]
{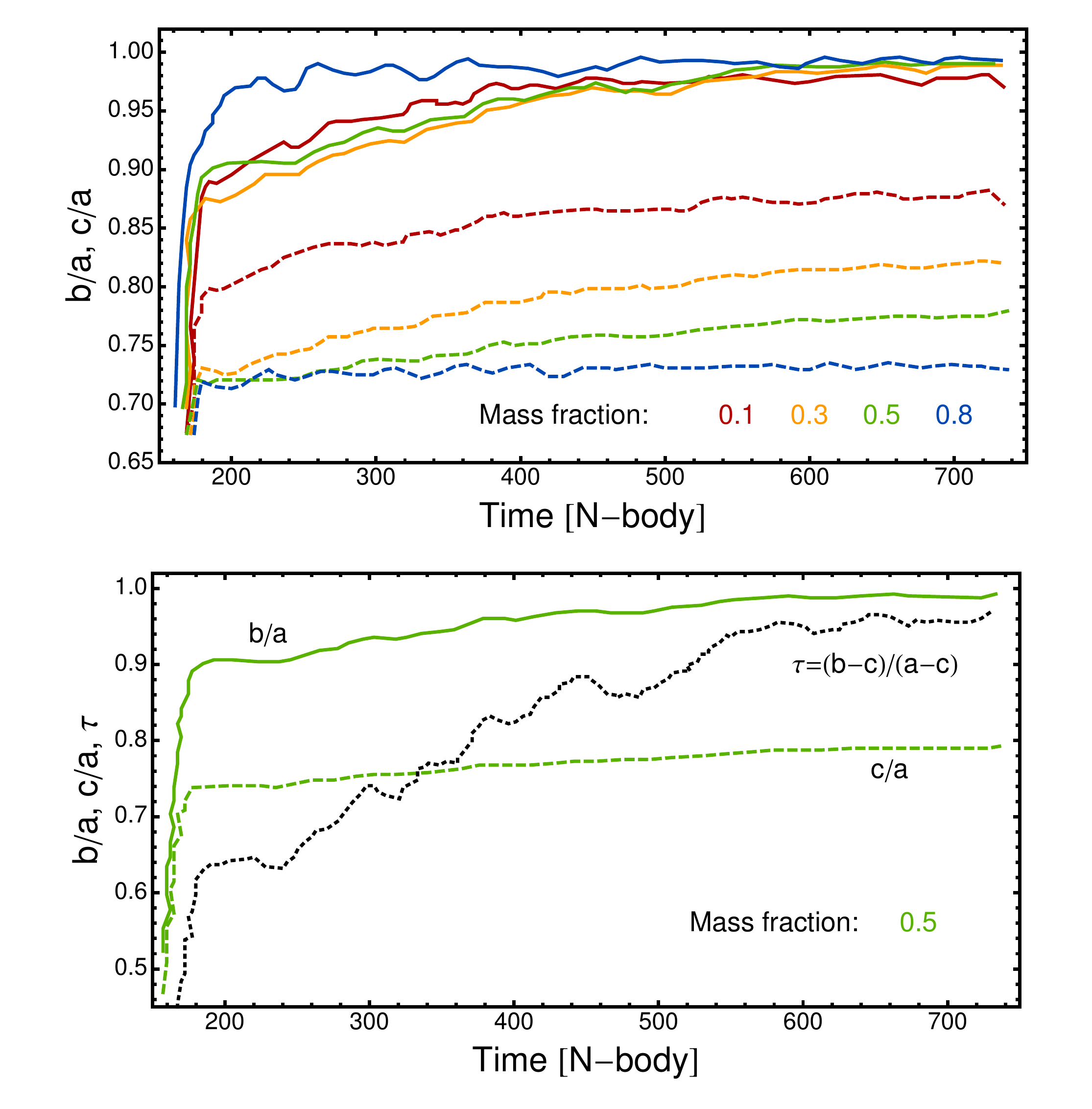}}
\caption{
Triaxiality of the resulting merged cluster for different mass
fractions (upper panel)
and the mass fraction 0.5.  
We calculate the semi-major axes of the ellipsoid of inertia a, b and c (where $a>b>c$)
according to four different mass fractions which, in turn, are distributed on
the basis of the amount of gravitational energy. The shorter the
distance to the center of the resulting cluster, the lower the mass fraction.
Displayed are $b/a$ (solid lines) and $c/a$ (dashed lines). The lower
panel shows the shape indicators for the mass fraction 0.5, together
with the 
evolution of the parameter $\tau$, an indicator for the triaxiality of the
system, which tends to one as time elapses; i.e. the system tends to be oblate.
The evolution of $\tau$ is similar for the rest of mass fractions
\label{fig.EvolTriax0p10p8}
}
\end{figure}

In Figure~\ref{fig.EvolTriax0p10p8} we show the role of the cluster symmetry
explicitly by depicting the evolution of the triaxiality of the cluster formed
as a result of the merger of the two clusters for our fiducial model in the case
of the double-cluster channel \citep[which is the reference model of][]{ASF06}.
After a merger which is the result of a parabolic orbit, the final system
is oblate rather than prolate; i.e. $a \sim b > c$, where $a$, $b$ and $c$
are the cluster axes. At the outskirts the resulting merged cluster is flatter
and at the center the binary of IMBHs makes it rather spherical.
\cite{Amaro-SeoaneEtAl09a}
addressed the single-cluster channel scenario after the formation of the IMBHs and used 
additional simulations to further
evolve the BBH. They used scattering experiments of three bodies including
relativistic precession to 1st post-Newtonian order, as well as radiation
reaction caused by GW, so that they did not have to integrate every single star
in the cluster to understand the posterior evolution of the BBH. In their work,
between the strong encounters, $a$ and $e$ of the BBH were evolved by resorting
to the quadrupolar formul{\ae} of \cite{Peters64}.

The BBH will have completely circularized when it reaches the
frequencies probed by Advanced LIGO
and the ET, because the emission of GWs takes over the dynamics
of the system.

\subsection{Death}

\begin{figure}
\resizebox{\hsize}{!}{\includegraphics[scale=1,clip]{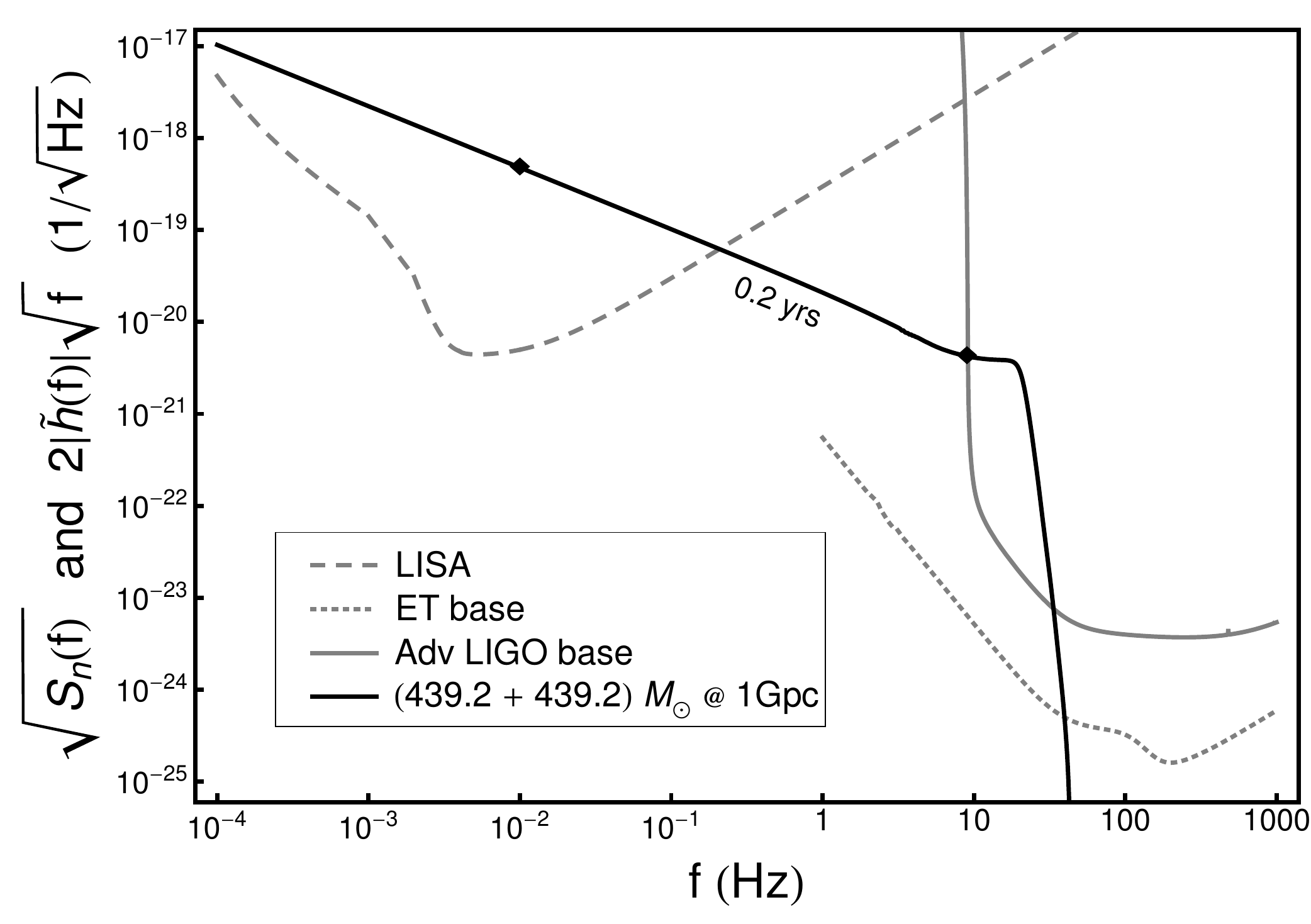}}
\caption{
Amplitude of the GW emitted by a system of two equal-mass IMBHs of total mass
$878.4 \, M_{\odot}$ at $100$~Mpc as seen by different GW
observatories.  Note that we have 
multiplied $\lvert \tilde{h}(f)\rvert$ by a factor $2\,\sqrt{f}$, with $f$ the
frequency of the system. This is required in order to be able to compare it
with the sensitivity curve of the different detectors (see section
\ref{sec:wfmodel} for more details). From left to right we depict the sensitivity
windows of the future space-borne LISA (dashed, gray curve), the Einstein
Telescope (dotted, gray curve) and Advanced LIGO (solid, gray line starting
sharply at 10~Hz). The strain of the BBH of IMBHs spends most of its
inspiral in the LISA band, whilst the ringdown and merger occur at higher
frequencies, only observable by ground-based detectors. Notably, the ET
captures an important extent of the inspiral as well as the whole ringdown and
merger. The averaged SNR produced by this system would be ${\rm
  SNR}_{\rm LISA} = 854$, ${\rm
  SNR}_{\rm ET} = 7044$ and ${\rm
  SNR}_{\rm AdvLIGO} = 450$. The BBH system spends approximately
0.2~yrs to go from 
$f=0.01\,\rm{Hz}$ 
(well into the LISA band) up to the lower cut-off
frequency of Advanced LIGO, 10~Hz. These two points are pinpointed on the plot
\label{fig.LISA}
}
\end{figure}

While the emission of GWs is present all the time from the very first moment in
which the BBH is formed, the amplitude and frequency of the waves is
initially so low that no
present or planned detector would be able to register any information from the
system. Only when the semi-major axis shrinks sufficiently does the frequency
increase enough to ``enter'' the LISA band, which we assume
starts at $10^{-4}$ Hz. The BBH then crosses the entire detector window
during its inspiral phase, as we can see in Figure~\ref{fig.LISA}.  We depict
the evolution of a BBH of mass $439.2 + 439.2\,M_{\odot}$. The reason for this
particular choice of masses is to give the reader a point of reference to
understand the whole picture. Recently \cite{AS10a} included the
effect of the rotation of the host cluster and addressed the dynamical
evolution of the global system with a BBH of that mass. 
The authors have shown that LISA will see the system of Figure~\ref{fig.LISA}
with a median SNR of few tens. The fact that the system merges outside its band
prevents LISA from observing the loudest part of the BBH coalescence. 
In order to follow the system at this early stage of its
evolution {\em in the LISA band}, a simple post-Newtonian approach suffices for
modeling the GW radiation. We are far enough from the highly relativistic
regime and only the inspiral phase of the BBH coalescence is visible to the
space antenna, with a rather low SNR compared to the posterior phases in the
evolution of the BBH.

As the binary system depicted in Figure~\ref{fig.LISA} leaves the LISA band and
enters the strong field regime, higher order post-Newtonian corrections and
eventually input from numerical relativity simulations need to be considered in
order to model the GW waveform.  Three reference frequencies in the evolution
of a compact BBH that approaches its merger are the innermost stable circular
orbit ($f_{\rm ISCO}$) of a test particle orbiting a Schwarzschild black hole,
the light-ring frequency ($f_{\rm LR}$) corresponding to the smallest unstable
orbit of a photon orbiting a Kerr black hole and the fundamental ringdown
frequency ($f_{\rm FRD}$) of the decay of the quasi-normal modes computed by
\cite{Berti:2005qd}.

\begin{figure}
\resizebox{\hsize}{!}{\includegraphics[scale=1,clip]
{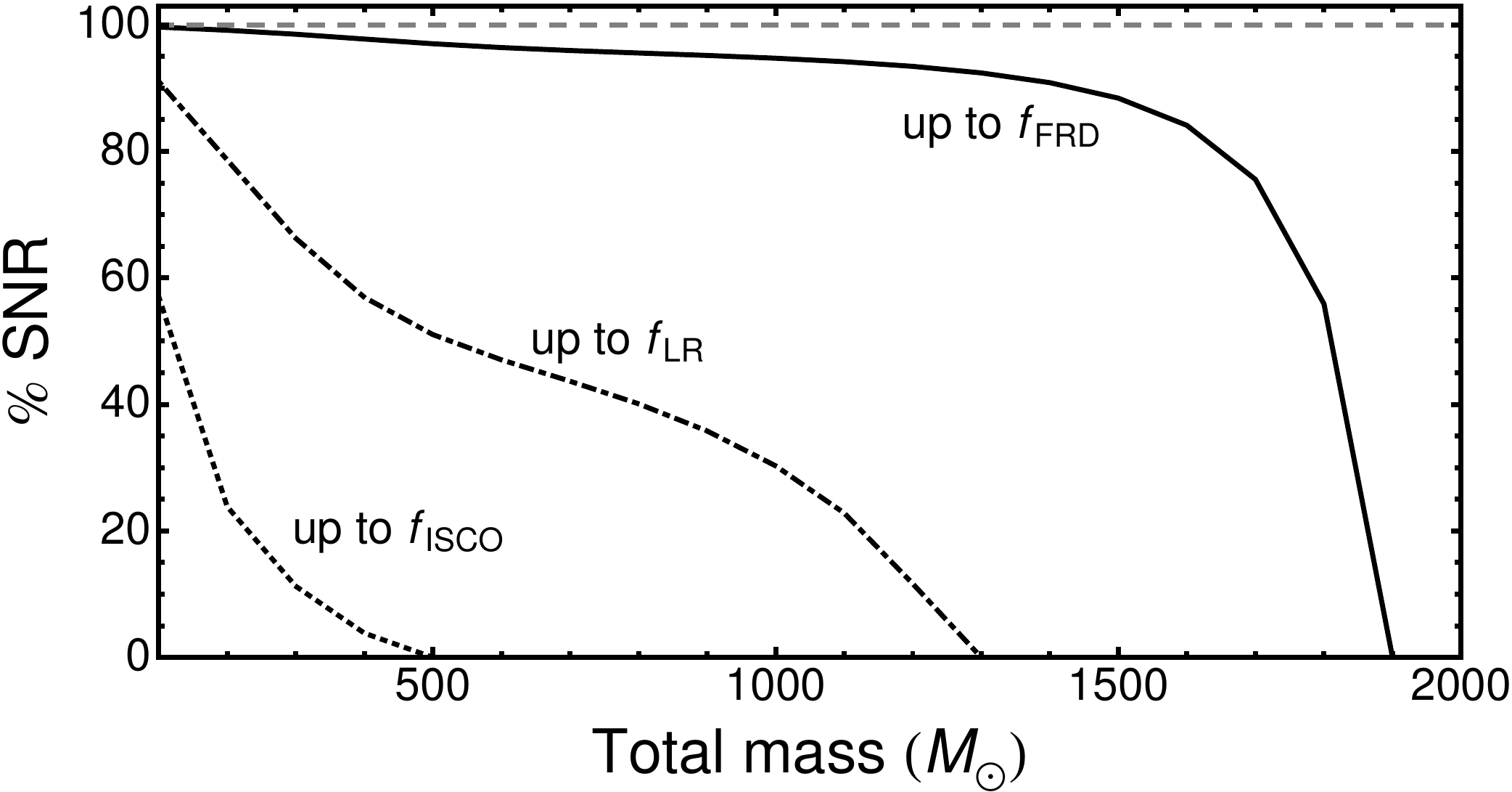}}
\caption{
Percentage of the total SNR produced by IMBH inspiralling signals cut
at the three reference frequencies $f_{\rm ISCO}, f_{\rm LR}$ and
$f_{\rm FRD}$. The SNRs have been calculated using the noise curve of
Advanced LIGO for signals placed at 100~Mpc of the detector starting
at 10~Hz and with a total mass between 100
and $2000\,M_{\odot}$. Whereas
the SNR computed up to $f_{\rm ISCO}$ constitutes more than 50\% of
the total SNR for systems with total mass below $100\,M_{\odot}$, it is the
merger and ringdown parts of the coalescence (after $f_{\rm LR}$ and
$f_{\rm FRD}$) that contribute most to the SNR as
the total mass of the system increases above a few hundreds of solar masses
\label{fig.percSNR}
}
\end{figure}

For the binary system shown in Figure~\ref{fig.LISA}, the values of these three
frequencies are $f_{\rm ISCO}|_{878.4 M_{\odot}}  \simeq 5 ~{\rm Hz}$, $f_{\rm
LR}|_{878.4 M_{\odot}} \simeq 14.2~{\rm Hz}$ and $f_{\rm FRD}|_{878.4
M_{\odot}} \simeq 21.4 ~{\rm Hz}$. Should such a binary exist at a distance of
100~Mpc (the choice for this number is based solely on the fact that is is
easily scalable), and if it were to be detected with Advanced LIGO, it should
produce a sky-averaged SNR of $\sim 450$, assuming a low frequency cut-off of
10~Hz. To that total SNR, the contribution of parts of the inspiral happening
before the system reaches the characteristic frequencies $f_{\rm ISCO}$,
$f_{\rm LR}$ and $f_{\rm FRD}$ would be 0\%, 37\% and 95\% respectively.
Figure~\ref{fig.percSNR} illustrates the same percentages for binaries with
total masses between 100 and $2000\,M_{\odot}$. It is immediately noticed that
for the binaries of IMBHs of interest to this study, most of the SNR that these
binaries will produce in Advanced LIGO comes from the last stages of the 
BBH coalescence.

We can estimate the time that the binary system takes to evolve from
$f=0.01\,\rm{Hz}$, a frequency where the BBH can be seen by LISA, to the lower
cut-off frequency of 10~Hz of Advanced LIGO or of 1~Hz of the ET.  A lower
order approximation based on the Newtonian quadrupole formula
\citep{Peters64} leads to the following expression for the evolution of
the frequency in terms of the chirp mass ${\cal{M}}= (m_1 m_2)^{3/5}M_{\rm
tot}^{-1/5}$ and frequency of the system

\begin{equation}
\frac{df}{dt}=\frac{96}{5}\pi^{8/3}{\cal{M}}^{5/3}f^{11/3}.
\label{eq:freqEvol}
\end{equation}

\noindent
We find a delay of only 0.2 yrs (80 days) for a BBH with total mass
$M=878.4\,M_{\odot}$ to go from 0.01 Hz to the beginning of the
Advanced LIGO band and almost similar numbers to the beginning of the ET
band (the evolution of the system is extraordinarily quick in the late
inspiral phase, which explains the fast evolution from 1 to 10~Hz). In
view of these figures, LISA could be used as an ``alarm'' to
prepare ground-based detectors to register the final coalescence in detail, the
death of the BBH as such, by adjusting their {\em sweet spots} (the most
sensitive part of the detector) to the particular BBH. The high
accuracy of parameter estimation during the inspiral phase of which
LISA is capable could be
combined with the information obtained from the large-SNR triggers that the BBH
merger and ringdown will produce in Advanced LIGO or ET and thereby achieve a more
complete characterization of the system. 

\begin{figure}
\resizebox{\hsize}{!}{\includegraphics[scale=1,clip]
{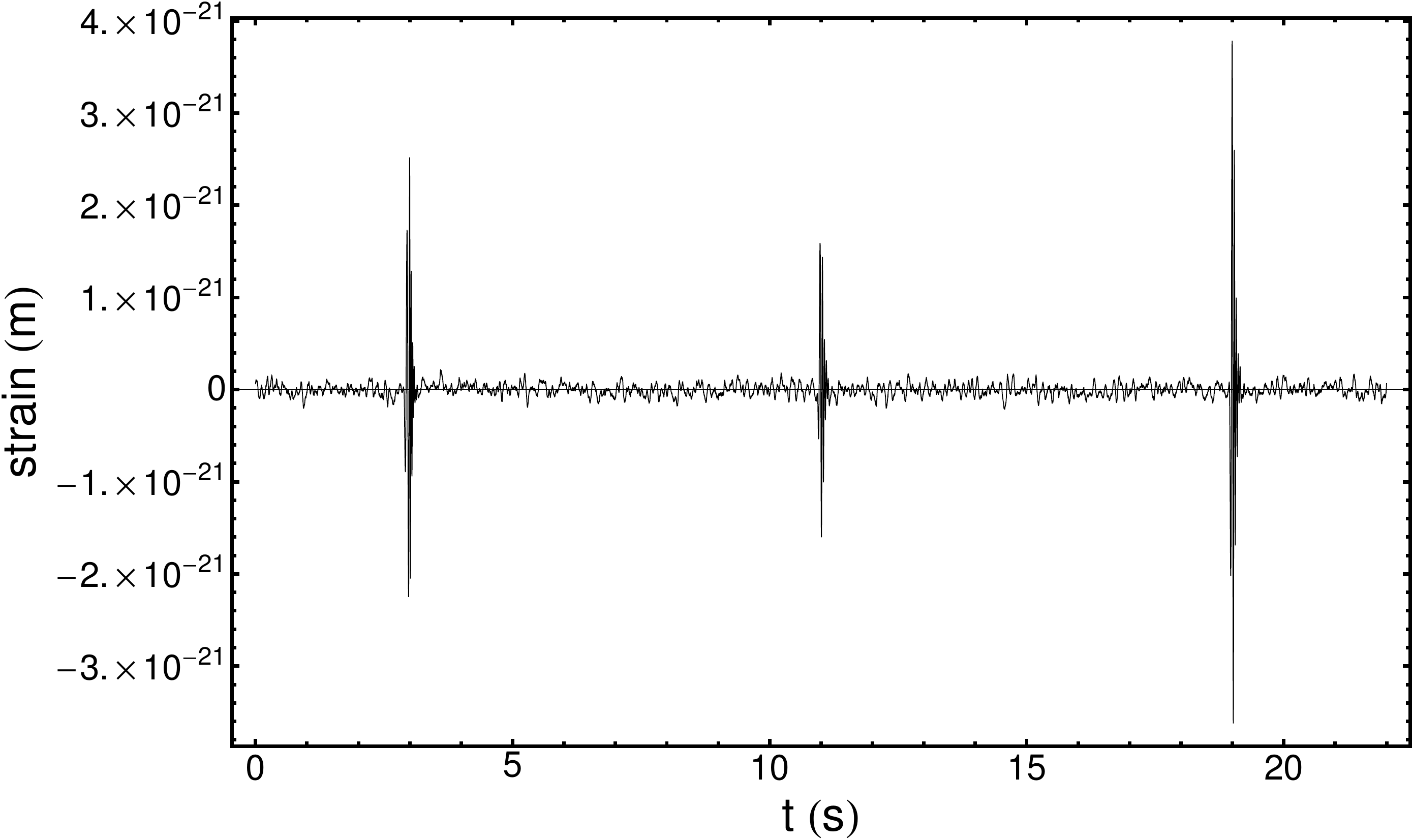}}
\caption{IMBH systems as will be seen in the time-domain output strain
  of the detector by the
  Advanced LIGO interferometer at the Livingston site. Three signals
  corresponding to equal-mass,
  non-spinning IMBH systems with total mass $400\,M_{\odot}\leq
  M_{\text{total}} \leq 700\, 
  M_{\odot}$ and random orientations and polarization angles have been
  placed at $1$~Gpc from the detector with a starting
  frequency of 10~Hz. The L1 interferometer strain has been modeled
  by Gaussian 
  noise colored with the design sensitivity curve expected for
  Advanced LIGO. Depending on their distance and orientation, the
  signals could be spotted by eye, which gives an intuitive idea of
  the kind of ``bright'' (in terms of GW emission) sources they are
\label{fig:IMBHsinNoise}
}
\end{figure}

\section{BBH waveform model and expected SNR}
\label{sec:wfmodel}

Accurate theoretical modeling of the gravitational radiation $h(t)$ emitted by
a BBH is key to improving BBH detectability and parameter estimation.
While post-Newtonian (PN) theory is
valid to model the early inspiral phase of the BBH evolution in the
LISA band, an exact 
description of the merger and ringdown stages is only possible via numerical
relativity (NR) calculations. Several approaches have been proposed to
match PN and NR in order to obtain a full waveform for the entire
coalescence, most notably for non-spinning
configurations
in~\cite{Buonanno:1998gg,BuonD00,Buonanno:2009qa,Ajith:2007qp,Ajith:2007kx}  
and recently introducing 
spins for non-precessing configurations in~\cite{Pan:2009wj,Ajith:2009bn}. 

\begin{figure}
\resizebox{\hsize}{!}{\includegraphics[scale=1,clip]
{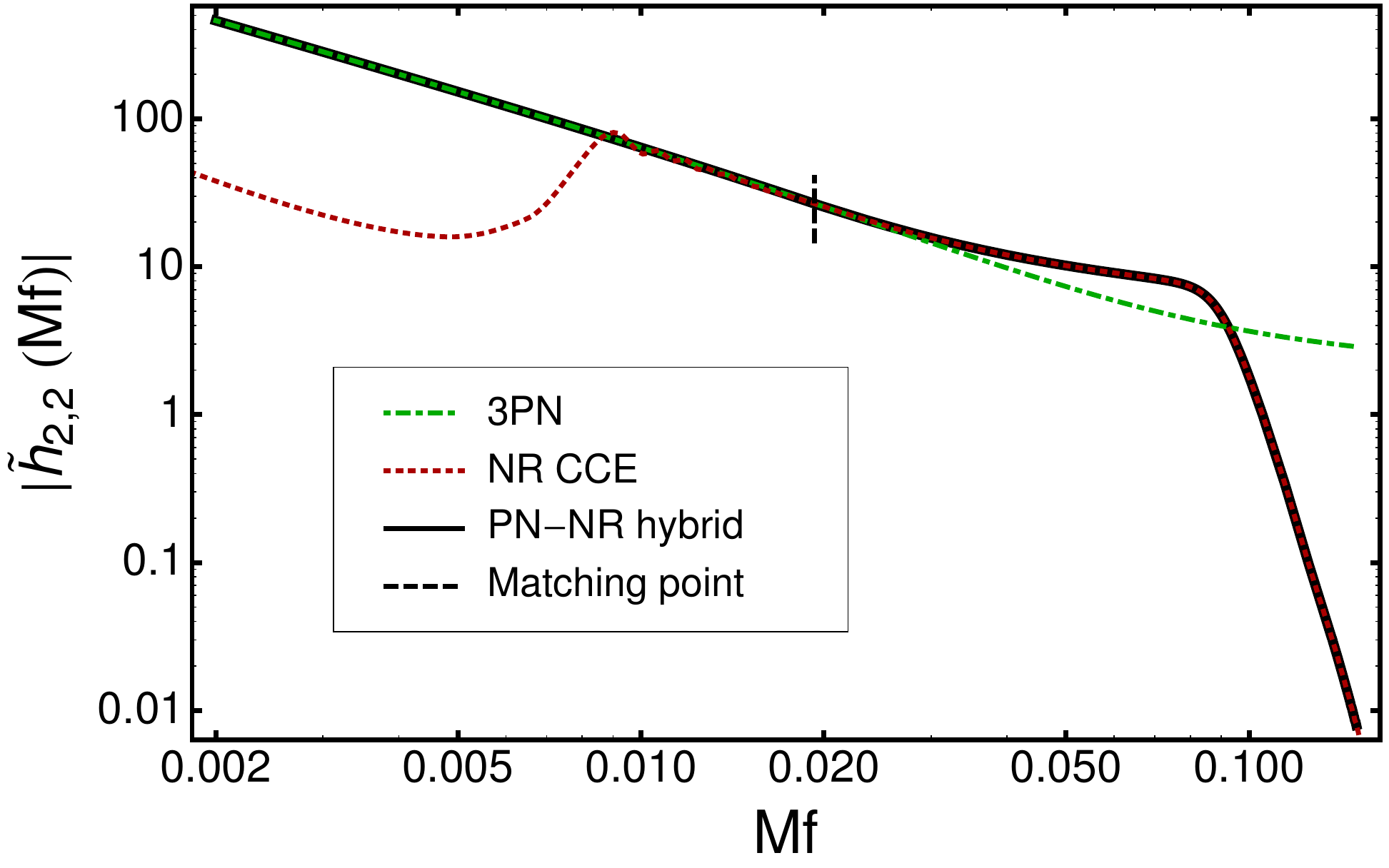}}
\caption{Waveform model employed in the SNR calculations of
  section~\ref{sec:wfmodel} for the equal-mass, non-spinning BBH
  scenario. The amplitude of the numerically simulated and
  $\mathscr{I}^+$-extrapolated $\ell=2, m=2$
  mode \citep{Reisswig:2009us} is attached to a PN
  calculation based on the stationary phase approximation that
  incorporates terms up to 3rd PN order. The amplitudes are stitched
  at a frequency $Mf \sim 0.02$ to produce a full
  IMR waveform. Note that the magnitudes
  displayed in the plot are dimensionless and can be scaled to account
  for different BBH masses
\label{fig:hybConstr}
}
\end{figure}

For the purpose of the SNR and horizon distance calculations shown in
this paper, we have chosen a 
new procedure for the construction of hybrid PN-NR waveforms in the
frequency domain 
proposed by~\cite{Santamaria:SpinTemp}. The construction procedure
developed for matching PN and NR data is sketched in 
Figure~\ref{fig:hybConstr}; further details on the fitting procedure
can be found 
in~\cite{Santamaria:SpinTemp}. In essence, the model consists of a 
phenomenological fit to hybrid PN+NR waveforms for spinning, non-precessing
BBH systems with comparable masses ($0.15\lesssim\eta<0.25$). This is
compatible with the simulations of \cite{GFR06}, which find pairs of
VMSs with mass ratios close to 1. The phenomenological
waveforms are parameterized by three physical parameters: the total
mass of the binary $M$, the symmetric mass ratio $\eta$ and the spin
parameter $\chi \equiv (1 + \delta) \chi_1/2 + (1-\delta) \chi_2/2$,
where $\delta 
\equiv (m_1-m_2)/M$, $\chi_i = S_i/m^2_i$ and $S_i$ is the spin
angular momentum of the $i$th black hole. Only the
dominant mode $\ell =2,\, m=2$ of the gravitational radiation enters
in the model; the effects of the higher modes, which become
increasingly significant as $\eta$ decreases, are neglected. For
comparable-mass scenarios, this restriction does not substantially
affect our results. The waveform model used in our calculations is
valid for binary systems with \emph{aligned} spins only, which represents
a first step towards the incorporation of the spins of the BHs in full
inspiral-merger-ringdown (IMR) models. A discussion including
precessing systems with 
arbitrary spins to this date still requires
work regarding availability of NR simulations and development
appropriate techniques to match precessing merger data with PN. Work
in this regard is ongoing. 

For the results presented in this article we choose to focus on three
configurations: (i) equal-mass, 
non-spinning, (ii) equal-mass, equal-value spins aligned with the
direction of the 
total angular momentum and magnitude $\chi=0.75$ and (iii) mass ratio
$1:3$ ($\eta=0.19$), non-spinning.  
Provided with the full IMR waveforms given by our model, 
we are interested in assessing the detectability of these systems by
current and future GW observatories. 
The SNR of a model waveform with respect to the output stream of the 
detector is the quantity typically quoted to signify the detectability of a
signal. The SNR $\rho$ produced by a GW signal $h(t)$ in a detector
can be computed as \citep[see e.g. ][]{Thorne300,Finn92}

\begin{equation}
\label{eq:snrDef}
  \rho^{2} \equiv 4\int_0^\infty \frac{\, \tilde{h}(f)\,
    \tilde{h}^\ast(f)} {S_n(f)}\, d f = \int_0^\infty
  \frac{\lvert 2\, \tilde{h}(f)\, \sqrt{f} \rvert^{2}}
  {S_n(f)}\, d\ln{f} ,
\end{equation}

\noindent 
where $\tilde{h}(f)$ is the Fourier transform of the strain $h(t)$. In the
last equation 
$S_n(f)$ represents the one-sided noise spectral
density of the detector.  

\begin{figure}
\resizebox{\hsize}{!}{\includegraphics[scale=1,clip]
{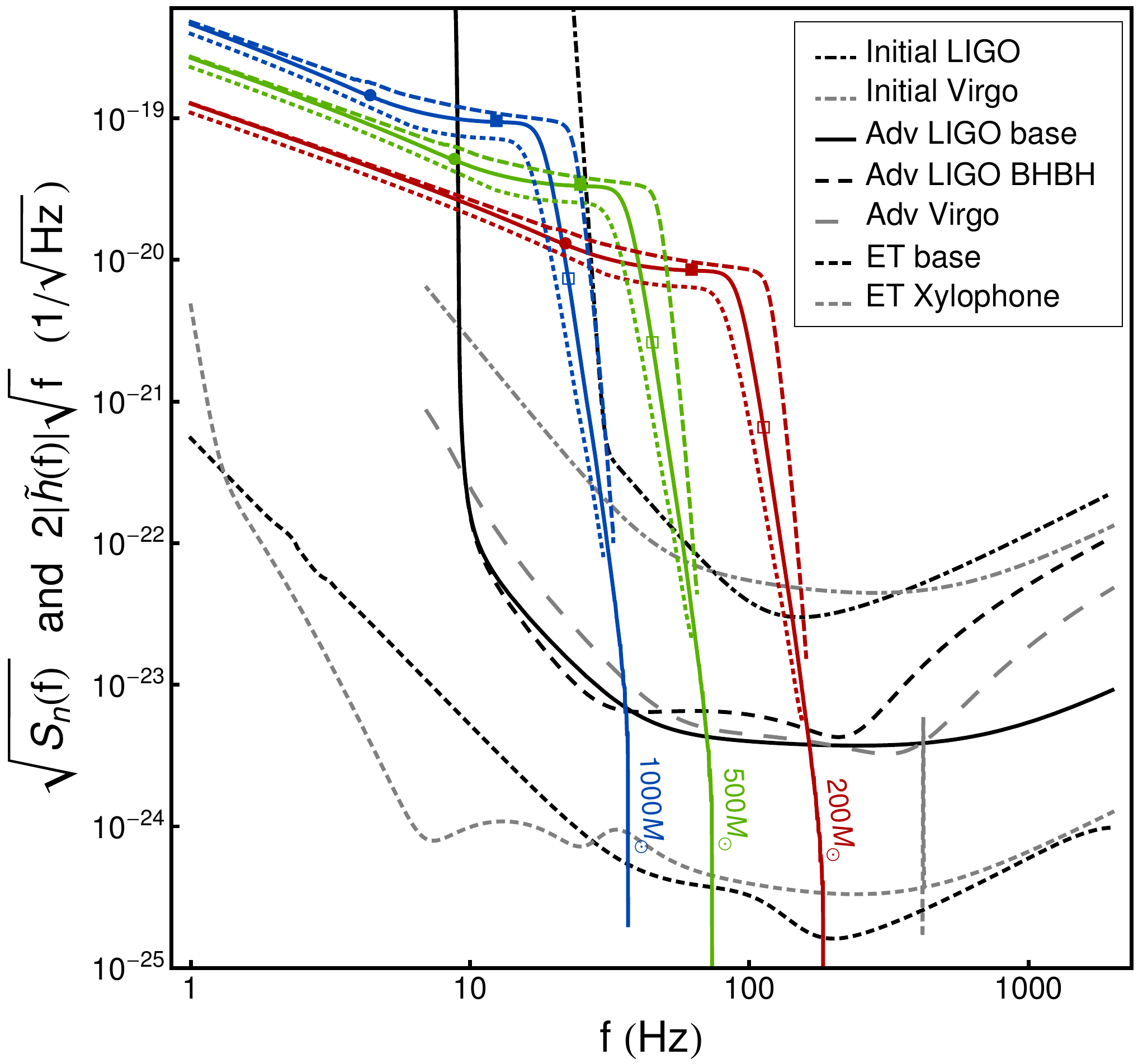}}
\caption{Hybrid waveform for three BBH configurations scaled to various IMBH
  masses. From top to bottom we show BBH systems with total mass
  $1000,  500$ and $200 M_\odot$, in blue, green and red
  respectively. Solid lines correspond to the equal-mass, non-spinning
  configuration (i), dashed lines to the equal-mass, $\chi=0.75$
  configuration (ii) and dotted lines to the non-spinning, $q=3$
  configuration (iii). The sources are optimally oriented and
  placed at 100~Mpc of the detectors. The symbols on top of
  configuration (i) mark various stages of the BBH
  evolution: solid circles 
  represent the ISCO frequency, squares the light ring frequency and
  open squares the Lorentzian ringdown frequency (corresponding
  to 1.2 times the fundamental ringdown frequency $f_{\rm FRD}$), when
  the BBH system has merged and the
  final BH is ringing 
  down. Currently operating and planned ground-based detectors are
  drawn as well: plotted are the sensitivity curves of initial LIGO
  and Virgo, two possible configurations for Advanced LIGO (zero
  detuning and $30-30 M_\odot$ BBH optimized), Advanced Virgo and the
  proposed Einstein telescope in both its broadband and
  xylophone configurations
\label{fig:SfvsH}
}
\end{figure}

Figure~\ref{fig:SfvsH} provides a graphical representation of the detectability
of several BBHs by current and future generations of ground-based GW
detectors. Displayed are the design sensitivity of current initial LIGO and
Virgo (sensitivities that have been met or approximately met during the S5/VSR1
data taking), the proposed noise curves of Advanced Virgo and two possible
configurations of Advanced LIGO (broadband or ``base'' and optimized for
$30-30\,M_{\odot}$ BBHs) and the designed noise budget for the Einstein Telescope
in its broadband, ``base'' and ``xylophone''~\citep{Hild:2009ns}
configurations. The hybrid 
waveforms constructed according to the model of
Fig.~\ref{fig:hybConstr} for the three chosen configurations has been
conveniently scaled  
to represent BBH systems with total mass $200,500$
and $1000\,M_\odot$. 

As the right-hand-side of Equation~\ref{eq:snrDef} suggests, 
plotting the quantity $2\, \lvert\tilde{h}(f)\rvert\, \sqrt{f}$ versus
$\sqrt{S_h(f)}$ allows for direct visual comparison of the importance of each
of the stages of the BBH coalescence. The three 
frequencies $f_{\rm ISCO},
f_{\rm LR}$ and the Lorentzian ringdown frequency  $f_{\rm LRD}=1.2
f_{\rm FRD}$ are 
marked on top of the configuration (i) waveforms with solid circles,
squares and open squares 
respectively. One can immediately appreciate that systems with total mass above
$500\,M_\odot$ fall almost completely below the 40~Hz ``seismic
wall'' of the 
initial LIGO detectors; however they will become very interesting
sources for the second generation of GW interferometers and the proposed
Einstein Telescope. Indeed, as we show in Section~\ref{sec:astro}, they
will also be seen by the future space-borne LISA. Additionally it is
easy to appreciate why the hang-up configuration (iii) will produce 
larger SNRs and therefore will be seen to further distances, for it
merges at higher frequencies with respect to a non-spinning
configuration with the same $\eta$. In
contrast, for equal spin values, the SNR decreases with smaller
symmetric mass ratios (compare configurations (i) and (iii)). 

\begin{figure}
\resizebox{\hsize}{!}{\includegraphics[scale=1,clip]
{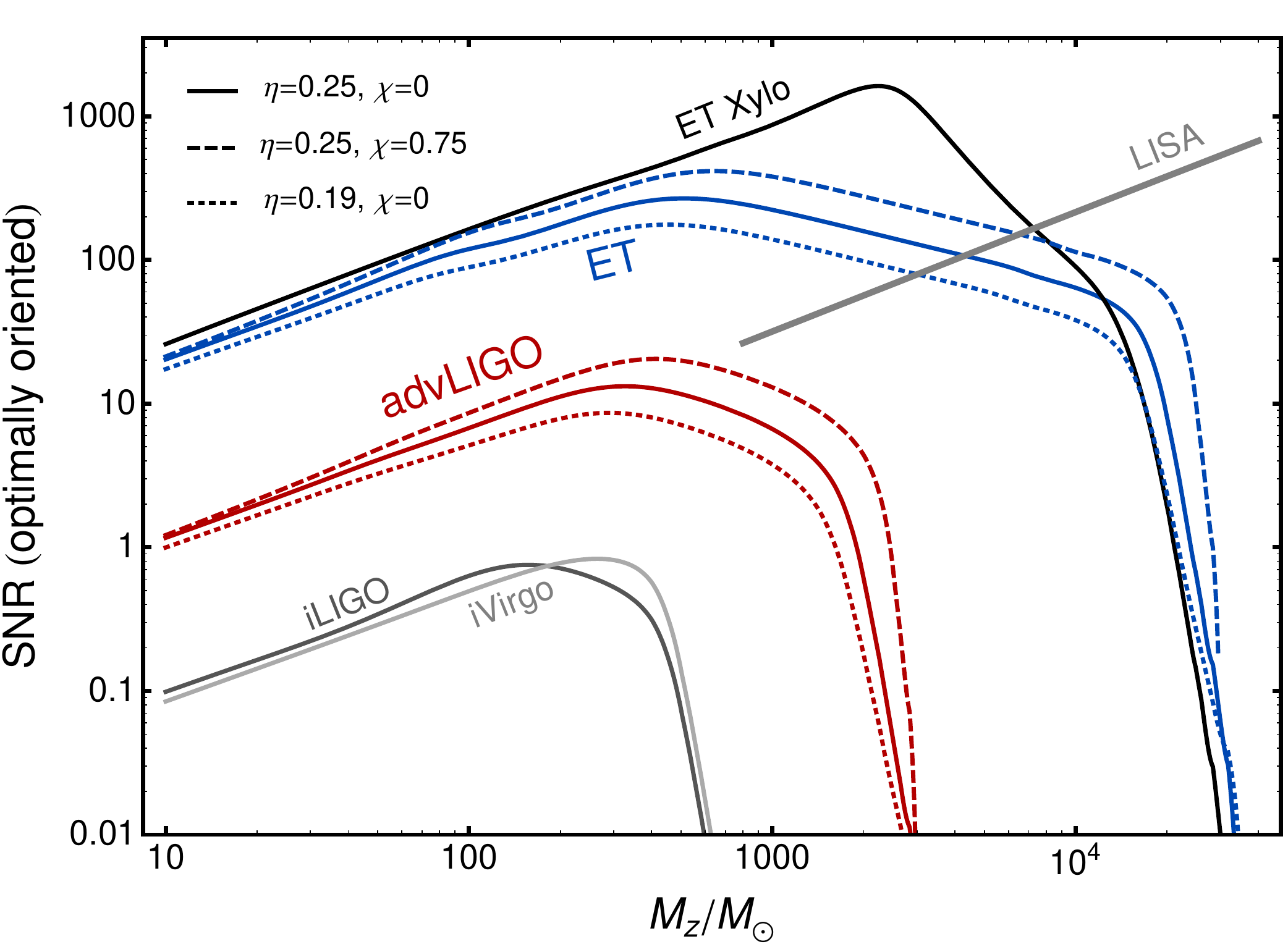}}
\caption{Signal-to-noise ratio as a function of the redshifted total mass of the
  BBH for the present and future generations of GW
  detectors and LISA. The sources are placed at a distance of
  6.68 Gpc ($z=1$) and the
  SNRs correspond to sources optimally oriented and located. Solid
  lines indicate SNRs for the equal-mass, non-spinning configuration
  (i); for Advanced LIGO and ET we have included the SNRs produced by
  configurations (ii) 
  and (iii) as well, indicated with dashed and dotted lines respectively
\label{fig:snrAll}
}
\end{figure}

In Figure~\ref{fig:snrAll} we compute the SNR expected for these
sources in each of the above-mentioned detectors as a function of the
redshifted total mass of the system $M_z=(1+z)\,M_{\rm BBH}$, for
optimally-oriented and -located sources at a distance of 6.68 Gpc
(z=1). We have cross-checked our SNR results with those computed
by~\cite{Boyle:2009dg} for initial LIGO at 100 Mpc, obtaining SNR
values within 1\% 
of those quoted by them. A direct comparison with~\cite{Gair:2009gr}
shows a disagreement of $\sim 30\%$ in the computed SNRs for Advanced
LIGO and the ET, which might be
attributed to the different waveforms and cosmological model
used. 

Unsurprisingly, the SNRs calculated for the third generation of ground-based
detectors surpass the expectations for initial and Advanced Virgo and
LIGO at all 
masses. At $z=1$, SNRs of the order of 0.1 are expected for current
LIGO and Virgo 
interferometers for binaries with total mass up to a few hundred
$M_\odot$. These values would scale to SNRs of $\sim 1$ at the closer distances
typically surveyed by the initial interferometers, which are of the
order of tens to few hundreds of Mpc. Signals with single-interferometer
SNRs below the commonly-used threshold of SNR=5.5 are most likely to be missed,
therefore IMBH binary coalescences are of limited interest for the
first-generation detectors --and even for their current,
\emph{enhanced} counterparts--, which are in turn most sensitive 
to neutron star binaries and stellar-mass back hole binaries.

Advanced LIGO and Virgo will be able to measure averaged SNRs
of the order of $1-20$ at $z=1$, with a maximal response to BBH systems
with total mass in the range of 300 to $800\,M_{\odot}$. 
Therefore, should
binaries of IMBHs exist in our neighboring Universe, the
second-generation of GW interferometers should be able to detect them,
for they will be loud enough to stand above the detector noise. As we
will see later, the non-negligible IMBH binary coalescence rate for
advanced LIGO indicates 
that these sources should be taken into account in 
future match-filtered searchers for inspiral binaries. 
Due to the moderate SNR of the signals, a potential detection might be
sufficient to confirm the 
existence of binaries of IMBHs but not enough to determine all its
parameters, such as mass ratio and spins, with sufficient accuracy;
For the advanced detectors, \cite{Cutler:1994ys} estimate an accuracy
of $\sim 1\%$ in the 
reduced mass $\mu = m_1 m_2/M$ using PN
templates  with negligible spins at SNR$\sim 10$. When they take the spins into
account, $\Delta \mu$ increases by $\sim 50$. It is expected that the
use of full IMR templates will improve these figures, but the final
accuracy will always be limited by the slightly-above-threshold SNRs
expected in Advanced LIGO and Virgo.
 
The Einstein Telescope will instead measure SNR values within the $10^2$ range
at $z \sim 1$, and
it is expected to be sensitive to binaries with total masses of the order of
$10^4\,M_{\odot}$, a significantly larger range than that surveyed by Advanced
LIGO and Virgo.  It is noticeable how the ET xylophone configuration increases
the detectability of binaries with masses above $1000\,M_{\odot}$ with respect
to the broadband ET configuration. This is due to its improved sensitivity
precisely at frequencies in the range of $1-30$~Hz, which is where systems of
mass above thousands of solar masses accumulate most of their SNR (see
Figure~\ref{fig:SfvsH}).  The large-SNR events that binaries of IMBHs would
produce in the ET are of great importance for astrophysics; the reason
being that as the SNR increases, the accuracy in parameter estimation
also does. In the limit of large SNR, the inverse of the Fisher
information matrix 
$\Gamma_{a\,b} = (h_{,\,a} | h_{,\,b})$ is an estimator of the errors in
the recovered parameters. At the relative large SNRs potentially
produced by IMBH binaries in the ET, the possibility of extracting
their mass ratios and spins to high accuracies would be revolutionary
for characterizing the IMBHs populations.   
 
As for LISA, BBHs with masses of hundreds of
solar masses will be seen with a moderate SNR \citep[see][for a detailed study of the parameter extraction]{AS10a} --~it is
only at masses above tens of thousands of solar masses that LISA will start
taking over the ground-based observatories. Although the space antenna
will be most sensitive to 
BBH binaries with masses in the range of $10^{\,6-7}\,M_{\odot}$, the possibility
that it can act as a complementary observatory for the Einstein Telescope for
IMBH binaries is very promising. Parameter accuracy studies for IMBHs in
LISA are already available using the inspiral part of the coalescence
\citep[including also non-negligible eccentricities;
see][]{AS10a}, 
and indicate that masses and sky positions will be recovered with a high
accuracy level. In order to complete the characterization of IMBHs with the
information given by the second and third generations of ground-based
detectors, a comprehensive study of parameter recovery taking the BBH
coalescence into account is very much desirable.

\section{Event Rates}
\label{sec:events}

\cite{Miller02} estimated for the first time the event rate for {\em
intermediate-mass} mergers of IMBHs (i.e. typically stellar black holes merging
with IMBHs) in clusters by calculating the luminosity distance for the
inspiral, merger and ringdown \citep{FlanaganHughes98} out to which these three
stages can be detected with an SNR larger than 10. In his approach,
the maximum distance for the detector was 3 Gpc ($z \sim 0.53$), with
no cosmological corrections. The event rate was calculated as
$R = \int (4/3)\,D(M)^3\,\nu(M)\,n_{\rm ng}\,f(M)\,dM$.
\noindent
In this equation $n_{\rm ng}$ is the number density of globular clusters, which
was taken to be $n_{\rm ng} \sim 8h^3/{\rm Mpc}^3$, as in the work of
\cite{PortegiesZwartMcMillan00}. The rate of coalescence of stellar-mass
compact objects with the IMBH is $\nu(M)$ and $f(M)=dN/dM$ is the mass
distribution of massive enough black holes in clusters. Obviously, $\int
f(M)\,dM = f_{\rm tot}< 1$.  \cite{Miller02} uses
the estimation of \cite{FlanaganHughes98} for it and finds that a few per year
should be detectable during the last phase of their inspiral.  Two years later,
\cite{Will04} revisited the problem using matched filtering for the
parameter estimation, an updated curve for the sensitivity of the detector and
restricted post-Newtonian waveforms to calculate an analytical
expression for the 
luminosity distance $D_{\rm L}$; his estimation is a detection
rate for binaries of about 1 per Myr
in a mass-range of $(10\,:\,100)\,M_{\odot}$.

\noindent 
Subsequently, \cite{FregeauEtAl06}
calculated the number of events that initial and Advanced LIGO (and LISA) could
see from the single-cluster channel. In their estimation, they assume that the
VMSs formed in the runaway scenario do not merge into one, but
evolve separately and eventually each form an individual IMBH, following the
numerical results of the Monte Carlo experiments of \citet{GFR06}. They derive a
generalized form of the event rate which can be summarized as follows 

\begin{equation}
R = \frac{d N_{\rm event}}{dt_0} = \int_0^{z_{\rm max}}\frac{d^2
  M_{\rm SF}}{dV_c dt_e} \, g_{\rm cl}\, g \,\frac{dt_e}{dt_0}\,
\frac{dV_c}{dz} \,\int_{M_{\rm cl,\,min}}^{M_{\rm
    cl,\,max}}\frac{dN^2_{\rm cl}}{dM_{\rm SF,\,cl} dM_{\rm cl}}\,dM_{\rm
  cl} \,dz.
\label{eq:ratesIntegral}
\end{equation}

\noindent
In this expression, $d^2  M_{\rm SF}/dV_c dt_e$ is the star
formation rate (SFR) per unit of comoving volume per unit of local
time; $g_{\rm cl}$ is the fraction of mass that goes into the massive
clusters of interest; $g$ is the fraction of massive clusters which
form IMBHs; $dt_e/dt_0 = (1+z)^{-1}$ is the relation between
local and observed time; $dV_c/dz$ is the change of comoving volume
with redshift; and $dN^2_{\rm cl}/dM_{\rm SF\,cl}\, dM_{\rm cl}$
is the distribution function of clusters over individual cluster mass
$M_{\rm cl}$ and total star-forming mass in clusters $M_{\rm
  SF\,cl}$; $z_{max}$ is the maximum redshift that the detector is
capable of observing. 

\begin{figure}
\resizebox{\hsize}{!}{\includegraphics[scale=1,clip]
{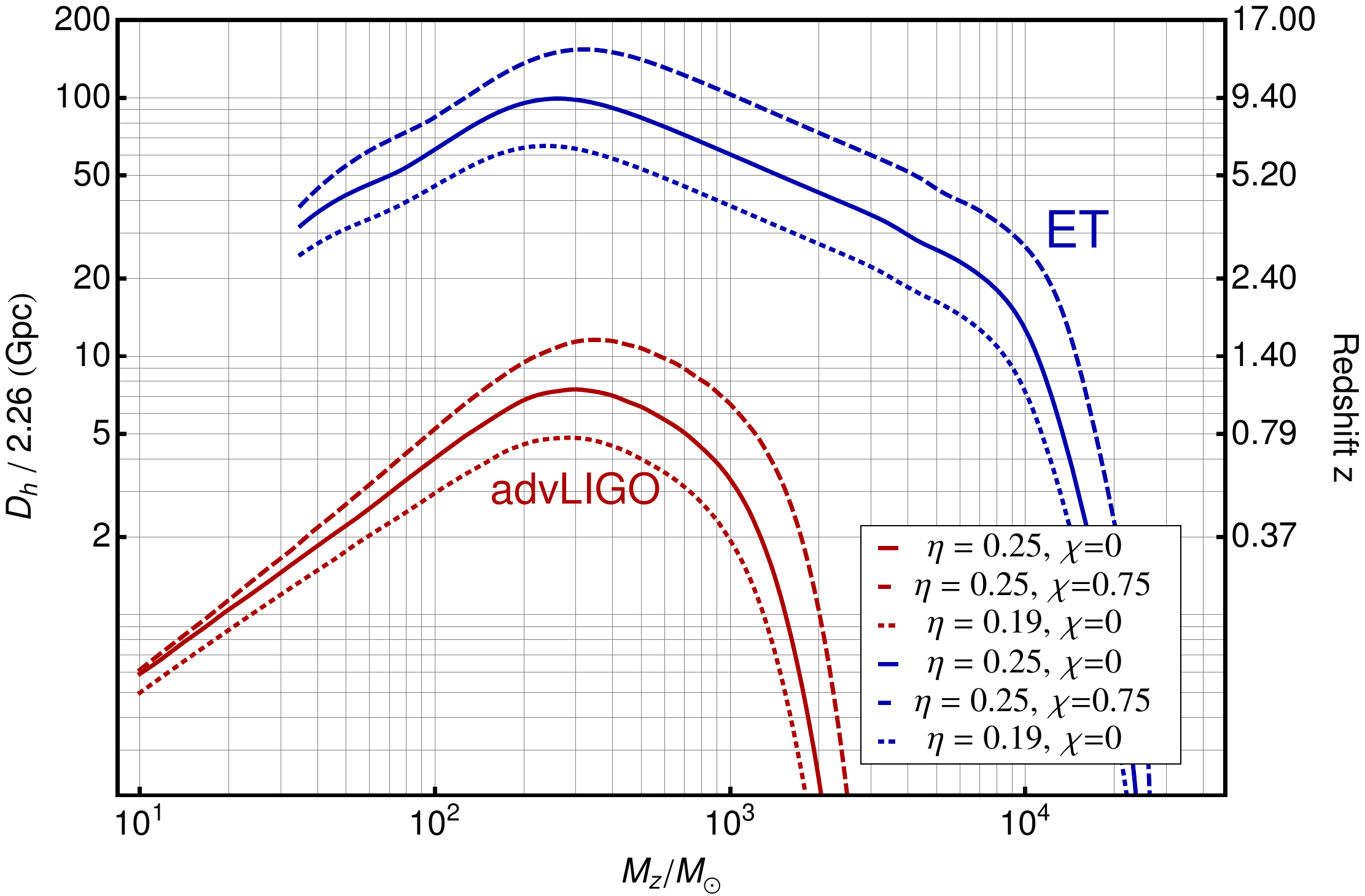}}
\caption{Orientation-averaged distance versus redshifted mass for three binary
  configurations
  obtained with the design sensitivity curves of Advanced LIGO and the Einstein
  Telescope. The solid, dashed and dotted lines corresponds to the
  configurations denoted in the text as (i), (ii) and (iii)
  respectively. Note the $\sim 40\%$ increase in reach given by the
  hang-up configuration with $\chi= 0.75$ with respect to the
  non-spinning case
\label{fig.HorDistPlot}
}
\end{figure}

In order to compute the integral above,
an estimation of the observable volume of each individual detector is required.
A commonly-used measure of the reach of a detector is the
horizon distance $D_h$, defined as the distance at which a detector
measures an ${\rm SNR}=8$ for an optimally-oriented and
optimally-located binary, i.e. an 
overhead, face-on orbit. Suboptimally located and 
oriented sources are detected with ${\rm SNR}=8$ at closer distances. 
However, in order to compute merger rates, an \emph{average} distance
over all possible 
orientations is more meaningful. The orientation-averaged
distance represents the cube root
of the total volume to which a detector is sensitive, assuming
uniformly-distributed sources, and is 2.26
times smaller than the horizon distance $D_h$~\citep{Finn:1992xs}. 
Moreover, at the distances that Advanced LIGO and the ET are expected to survey,
a certain cosmological model needs to be assumed. We adopt the standard
$\Lambda$CDM universe with parameters given by the first five years of
the WMAP sky survey~\citep{Hinshaw:2008kr}. These are $\Omega_\Lambda =
0.73$, $\Omega_b = 0.046$, $\Omega_c = 0.23$, $H_0 = 70.5\,{\rm
  km\,s}^{-1}{\rm Mpc}^{-1}$ and $t_0 = 13.72\,{\rm Gyr}$. 
Using the full IMR waveforms described in
Sec.~\ref{sec:wfmodel} and the corresponding redshift function $z(d)$ for the
$\Lambda$CDM model, we compute the orientation-averaged distance
$D_h/2.26$ 
for non-spinning systems with symmetric mass ratio $\eta = 0.25, 0.19$
and for an equal-mass system with spins aligned in
the direction of the angular moment and total spin $\chi=0.75$
(hang-up configuration). Computing $D_h (M_z)$ is essentially
equivalent to inverting Eq.~\ref{eq:snrDef} for $\rho(d,M(d))$ for $\rho=8$
using the relation $z(d)$ given by the cosmological model. The
results for $D_h (M_z)$ for our three configurations can be seen in 
Fig.~\ref{fig.HorDistPlot} for Advanced LIGO and ET.  

\begin{figure}
\resizebox{\hsize}{!}{\includegraphics[scale=1,clip]
{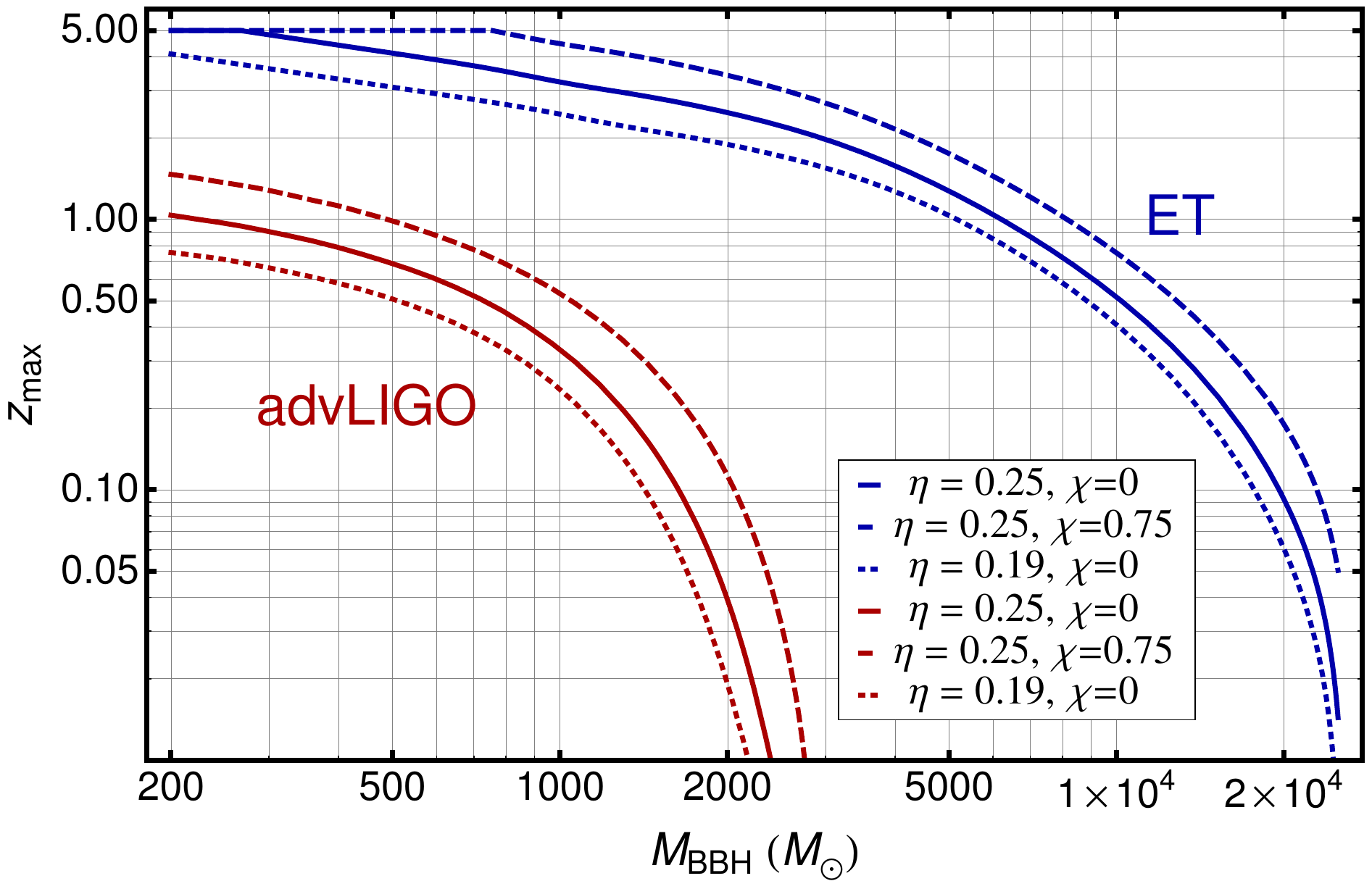}}
\caption{ 
Orientation-averaged distance expressed in terms of maximum redshift $z_{\rm
  max}$ for the Advanced LIGO and the ET detectors up
to $z = 5$ versus intrinsic total mass of the BBH. Red lines are for
Advanced LIGO and blue are for the ET. The solid, dotted, dashed
curves represent the same physical configurations displayed in figure
~\ref{fig.HorDistPlot} 
\label{fig:zMax}
}
\end{figure}

The limit for the integral in $z$ is thus given by the observable volume
of the detector of interest, quantified using the orientation-averaged distance.
The maximum redshift $z_{\rm max}$ is in turn a function of
the mass of the binary system, its configuration --mass ratio, spin--
and the particular waveform model used in the calculation of the
horizon distance. We present our values for $z_{\rm max}$ in
Fig.~\ref{fig:zMax}, which essentially 
shows the same data as 
Fig.~\ref{fig.HorDistPlot} but expressed in terms of redshift versus total
mass of the binary. 
As we see in Fig.~\ref{fig.HorDistPlot}, the maximum values for the
orientation-averaged distance for ET are as large as $z
\sim 10$ for configuration (ii) at $M_z \sim 300M_\odot$.  This implies that
the ET will be able to probe the 
different proposed scenarios to produce the first generation of black hole
seeds, as pointed out by \cite{SGMV09}. 
However, at these large cosmological distances, the stellar formation
rate is unknown and the validity of the rate integral cannot be stated.
We therefore set a maximum cutoff value of $z_{\rm max} = 5$ in the
computation of Eq.~\ref{eq:integralRates}. The final
value of $z_{\rm max} (M_{\rm BBH})$ that we have used in the computation of
the rates is show in figure~\ref{fig:zMax} for our three particular
physical configurations. 

Regarding the term in the $dM_{\rm cl}$
integral in~\ref{eq:integralRates}, three different parameterizations
of the SFR are available in the 
literature, i.e. $R_{\rm SF1,2,3}$ as summarized by
equations 4, 5 and 6 of~\cite{PorcianiMadau01}. The three models are similar
for distances up to $z \sim 2$, where SFR peaks, differing from there
on. For the 
results shown here, we have compared the three of them. As for the
distribution of cluster masses, the factor can be approximated as

\begin{equation}
\frac{dN^2_{\rm cl}}{dM_{\rm SF,\,cl} dM_{\rm cl}} = \frac{f(M_{\rm
    cl})}{\int M_{\rm cl}\,f(M_{\rm cl})\,dM_{\rm cl}}.
\end{equation}

\noindent In order to compute the integral in the denominator, we take $dN_{\rm
  cl}/dM_{\rm cl} \propto 1/M^2_{\rm cl}$, following the 
power law form observed by~\cite{ZhangFall99} for young star clusters
in the Antenn{\ae}. The validity of 
assuming the same law for the larger volume of the Universe surveyed by
Advanced LIGO or the ET is, however, a generalization not based on direct
observations. Thus, we should take this premise with care. By assuming an
efficiency factor of $f_{\rm GC}\sim 2\times10^{-3}$, based in the results of
\cite{GFR06}, we can set the values for the limits in $M_{\rm cl} =
f_{\rm GC} / M_{\rm BBH}$
according to the masses of the IMBH binaries of interest. In our case,
taking the standard definition of IMBH into account, 
this is $M_{\rm cl,\,max}/M_{\rm cl,\,min} = M_{\rm BBH,\,max}/M_{\rm
  BBH,\,min} = 2 \times 10^4 M_\odot / 2\times 10^2 M_\odot$.
The integral can now be expressed as

\begin{equation}
R = \frac{f_{\rm GC}}{\ln(M_{\rm cl,\,max}/M_{\rm
    cl,\,min})} \, g \, g_{\rm cl} \int_{M_{\rm BBH,\,min}}^{M_{\rm
    BBH,\,max}}\frac{dM_{\rm BBH}}{M^2_{\rm BBH}} \int_0^{z_{\rm
    max}(M_{\rm BBH})} SFR_i(z)\,  F(z)\,\frac{1}{(1+z)}\, \frac{dV_c}{dz} dz,
\label{eq:integralRates}
\end{equation}

\noindent
where $M_{\rm BBH,\,max\,(min)}$ is the range of total mass of the BBH
that we are considering, $SFR_i\,(z), \, i = 1,2,3$ is any of the three
considered stellar formation rates of~\cite{PorcianiMadau01} and

\begin{equation}
F(z) = \frac{\sqrt{\Omega_M(1+z)^3 + \Omega_k(1+z)^2 + \Omega_\Lambda}}{(1+z)^{3/2}}
\end{equation}

\noindent
is the factor that relates the stellar formation rate function in different
cosmologies with respect to the Einstein-de Sitter Universe. 

Regarding the values of the terms $g$ and $g_{\rm cl}$, it is
unfortunately very little what we know about the initial cluster
conditions required to form an IMBH binary. These factors have 
therefore large uncertainties. Following the existing literature, we
leave the fraction of massive clusters that form IMBHs, $g$, as a
parameter and set it to $0.1$ as an 
example. Nevertheless, as proven in the simulations of 
\cite{FreitagRB06}, it could be {\em as large as $0.5$}.  As for
the fraction of mass going into massive clusters, $g_{\rm cl}$,
previous works have also tentatively set it to $0.1$. 
Nevertheless, the results of~\cite{McLaughlin99} seem to indicate that
this value might be too optimistic. 
\cite{McLaughlin99} estimates an empirical cluster formation efficiency by mass 
of $\epsilon \equiv M^{\rm init}_{GCs}/M^{\rm init}_{\rm GAS}$, where $M^{\rm
init}_{GCs}$ is the initial globular cluster population and $M^{\rm init}_{\rm
GAS}$ is the initial reservoir available in the protogalaxy.
He then infers a universal value of $\epsilon \simeq 0.25\%$ after
evaluating a sample of 97 giant ellipticals, brightest cluster galaxies and
faint dwarfs. He finds identical results for the Population II spheroid of the
Milky Way. Therefore we have chosen to set our $g_{\rm cl}$ to
$1/400$, noting that, in any case, the final rates scale trivially
with both $g$ and $g_{\rm cl}$. 

We have evaluated the integral~\ref{eq:integralRates} for the three stellar
formation rates we consider and the three binary configurations (i),
(ii) and (iii) for Advanced LIGO and the ET. Our results
are summarized in table~\ref{tab:eventRates}. We find that the
event rate does not depend strongly on the assumed stellar formation rate; the
differences are of the order of $\sim 5\%$ for Advanced LIGO and $\sim
30\%$ for the ET. In the case of Advanced LIGO, the insensitivity to
the stellar formation model is easily explained by the fact that the
three models of~\cite{PorcianiMadau01} 
are very similar until $z\sim2$; our integration in
$z$ stops at $z\sim 1$ (see red curves in Fig.~\ref{fig:zMax}),
therefore ignoring contributions at higher $z$. For the
ET, the differences among $SFR_i\,(z), \, i = 1,2,3$ between $z \sim 2$
and 5 are very much attenuated by the rapid
decrease in comoving volume at high redshift. In order to assess the
effect of stopping the integration at $z=5$, we have also computed
the rates for the ET without imposing this condition. The differences
amount to $\sim 10\%$, therefore we conclude that the rates
estimations are not very sensitive to very high-$z$ SFRs. Nonetheless, we
should stress again that the lack of stellar formation data at
these large cosmological distances makes assumptions at $z>5$
largely speculative and thus we restrict to presenting the rates calculated
with $z_{\rm max}$ clamped at a value of 5. 
Given the inherent uncertainties present in our 
approach, which can amount to a few orders of magnitude, the
factors associated to the choice of SFR do not represent a major
source of error.    
We therefore quote the results found for $SFR_2(z)$ only,
corresponding to the star formation rate that increases up to $z \sim
2$ and keeps constant afterwards. 

\begin{table}[ht!]
\caption[Event rates for IMBH binaries formed in the single-cluster
channel potentially observable by
Advanced LIGO and the ET per year]{Event rates formed in the single-cluster
channel for IMBH binaries potentially observable by
Advanced LIGO and the ET per year. We take $g=0.1$, $g_{\rm cl} = 1/400$
as standard scaling values\label{tab:eventRates}} 
\begin{center}
\begin{tabular}{cccc}
\toprule
Detector & Configuration &
$R\left[\left(\displaystyle{\frac{g}{0.1}}\right)\,
\left(\displaystyle{\frac{g_{\rm cl}}{1/400}}\right)\,{\rm yr}^{-1}\right]$ \\
\midrule
Advanced LIGO & $\eta = 0.25,\, \chi = 0$ & $0.85$ \\
& $\eta = 0.25,\, \chi = 0.75$ & $2.36$ \\
& $\eta = 0.19,\, \chi = 0$ & $0.31$ \\
\midrule
Einstein Telescope & $\eta = 0.25,\, \chi = 0$ & $21.5$ \\
& $\eta = 0.25,\, \chi = 0.75$ & $24.0$ \\
& $\eta = 0.19,\, \chi = 0$ & $16.8$ \\
\bottomrule
\end{tabular}
\end{center}
\end{table}

The event rates do, however, strongly depend on the spins and mass
ratio of the binary. 
As expected, ``loud'' configurations like the hang-up case increase the event
rate by a factor of $\sim 3$ in the case of Advanced LIGO. This effect
is more notable the larger the total spin of the binary is; the
correct determination of the spin distribution of IMBH binary systems
would be extremely 
useful to further quantify the impact on the total rates. Smaller mass ratios
decrease the event rates, the difference between $\eta = 0.25$ and $\eta = 0.19$
being also approximately of a factor of 3.  The differences are not so
extreme in the case of the 
ET, due to the fact that we are cutting off $z_{\rm max}$ at a value of $5$
and, thus, neglecting contributions at higher redshift which might
amount to another $\sim 10 \%$, as discussed above. Even so, it is
evident that the expected rates increase with the 
total spin $\chi$ and the symmetric mass ratio $\eta$. The exact
computation of the 
total rates for all possible physical configurations of the binary
would necessitate further integration on the mass ratios and spins of
the system. At present it is not clear what that distribution might be,
therefore we simply summarize our results for all configurations under
consideration in table~\ref{tab:eventRates}. We find event rates of
the order of $\sim 1\,(g/0.1)\,(g_{\rm cl}/1/400)\, {\rm yr}^{-1}$ for Advanced
LIGO and of the order of $\sim 20\,(g/0.1)\,(g_{\rm cl}/1/400)\, {\rm yr}^{-1}$
for the ET. These rates assume formation of
IMBH binaries in the single-cluster channel. 

For Advanced LIGO and a non-spinning 
configuration,~\cite{FregeauEtAl06} found a expected detection rate of
$10 \,(g/0.1)\,(g_{\rm cl}/0.1)\, {\rm yr}^{-1}$ ($0.25\, {\rm
  yr}^{-1}$ in our units), adopting a 
reach of 2 Gpc ($z_{\rm max} \sim 0.37$). This estimation, calculated assuming a
ringdown-only search,
underestimates the volume of the Universe that Advanced LIGO would
observe using full IMR templates, as we show in
Fig.~\ref{fig.HorDistPlot}. Our results for non-spinning systems are compatible
with this observation; we quote rates larger than
those of~\cite{FregeauEtAl06} by a factor of 3 but otherwise of a
compatible order of magnitude.  
In the case of the ET,~\cite{Gair:2009gr} quote
a rate of $\sim 500 \,(g/0.1)\,(g_{\rm cl}/0.1)\, {\rm yr}^{-1}$
($\sim 13 \, {\rm yr}^{-1}$ in our units) for non-spinning
configurations, a value of the same order of
magnitude than the one found by us. The fact that~\cite{Gair:2009gr}
use a different family of 
waveforms to model the coalescence and a fitted formula for the
the averaged distance based on EOBNR waveforms, together with the slightly
different values in the integration limits explain the discrepancies
in the exact figures.   
Therefore, our new results for the rates of IMBH
binary coalescence for
non-spinning systems agree reasonably well with previous works that used
similar methods but different waveform and detector models. In
addition, we have
now quantified the effect that the spins of the BHs will have on the expected
rates and have 
found it to be of a factor $\sim 3$ for total spins of the binary as
high as $\chi = 0.75$

Thus far we have concentrated on the single-cluster channel scenario.
\cite{ASF06} give a prescription to calculate an
estimate of the event rates for the double-cluster channel. This was
based in the fact that 
the only difference, in terms of the event calculation, between both
astrophysical scenarios  
involves, firstly, that in the double-cluster channel there is
one single IMBH in one cluster and, secondly, that these two clusters
must collide so that the IMBHs form a BBH when they sink to the center
due to dynamical friction.
As explained in section 4 of \cite{ASF06}, the connection between the event
rate estimation of the two channels is
\begin{equation}
\Gamma^{\rm \,doub} = P_{\rm merg}\,g\,\Gamma^{\rm \,sing},
\end{equation}

\noindent
where $\Gamma^{\rm doub}$ is the event rate of the double-cluster channel, 
$\Gamma^{\rm sing}$ of the single-channel and $P_{\rm merg}$ is the
probability for two clusters to collide in the scenario of
\cite{ASF06}. They find that the unknown parameter $P_{\rm merg}$
could have values in the range $P_{\rm merg} \in [0.1,1]$. The total
event rate assuming that both formation channels are possible would be
\begin{equation}
\Gamma^{\rm \,tot} = \Gamma^{\rm \,sing} \left(\frac{g}{0.1}\right)
\left(\frac{g_{\rm cl}}{1/400}\right) \left(1+ P_{\rm
    merg}\,\left(\frac{g}{0.1}\right)\right), 
\end{equation}
Assigning parameter $P_{\rm merg}$ its pessimistic and optimistic
limits, we can compute the lower and upper limit of the event rates for
Advanced LIGO and the ET, assuming
contributions of the two channels. Taking into account the
double-cluster channel increases the rates upper limit by a factor of
2. For instance, for the equal-mass, non-spinning
case, the values are
\begin{align}
\Gamma_{\rm Adv.\,LIGO}^{\rm total} & \in [(0)\,0.94,\,1.7]\,\left(\frac{g}{0.1}\right)
\left(\frac{g_{\rm cl}}{1/400}\right)\,{\rm yr}^{-1}\\
\Gamma_{\rm ET}^{\rm total} & \in [(0)\,23.7,\,43.0]\,\left(\frac{g}{0.1}\right)
\left(\frac{g_{\rm cl}}{1/400}\right)\,{\rm yr}^{-1}
\end{align}

\noindent
Note that the results are quoted in terms of $g$ and $g_{\rm cl}$; the
already-mentioned uncertainty in their values could increase
these rates by at least one order of magnitude. These event
rates are encouraging to address the problem of detection
and characterization of systems of IMBH binaries with ground-based GW
observatories.  On the other
hand, one should bear in mind that the existence of IMBHs altogether has not
yet been corroborated, so that the pessimistic estimate is still somewhat
``optimistic''.  This is why we have added a $(0)$ in the previous rates as the
absolute lower limit.

\section{Conclusions}
\label{sec:concl}

Even though we do not have any evidence of IMBHs so far, a number of
theoretical works have addressed their formation in dense stellar clusters. If
we were to follow the same techniques that have led us to discover the SMBH in
our own Galaxy, we would need the Very Large Telescope interferometer and
next-generation instruments, such as the VSI or GRAVITY, which should be
operative in the next $\sim 10$ yrs. An alternative or even complementary way
of discovering IMBHs is via their emission of GWs when they are in a BBH
system.  

The identification and characterization of these systems rely on accurate
waveform modeling of their GW emission, which has been made possible by
the success of numerical relativity in simulating the last orbits of the BBH
coalescence and the coupling of these results to analytical post-Newtonian
calculations of the inspiral phase. We use a PN-NR hybrid waveform
model of the BBH coalescence based on a construction procedure in the frequency
domain \citep[see][for details]{Santamaria:SpinTemp}. 

Using this hybrid waveform, we have estimated the
SNR corresponding to the current and Advanced LIGO and Virgo detectors, the
proposed ET and the space-based LISA at a distance of $z=1$, i.e. 6.68
Gpc. The results indicate that IMBH binaries will produce
SNRs sufficient for detection in advanced LIGO and Virgo and notably larger
SNRs in the ET, thus making them interesting
sources to follow up on.
Eventual observations of IMBH binaries with future ground-based
detectors could be 
complementary to those of LISA, which is expected to detect these systems with
moderate SNRs and to be more sensitive to SMBH binaries instead. More
remarkably, in principle, if LISA and the ET are operative at the same time,
they could complement each other and be used to track a particular
event. 

Furthermore, we have revisited the event rate of BBHs for
various detectors and find encouraging results, within the inherent
uncertainties of the approach. Our estimations are consistent with
previous works, and additionally we have quantified the effect of the
total spin of the binary in the expected event rates.
We have estimated the distance to which Advanced LIGO and the ET will
be able to see binaries of IMBHs. This quantity depends strongly on
the mass ratio and spins of the binary. For Advanced LIGO, equal-mass,
non-spinning 
configurations of observed
total mass $\sim 200$--$700\,M_\odot$ can be seen up to $z \sim 0.8$. If
the spins are aligned with the total angular momentum and significant
($\chi_{1,2} \sim 0.75$), the reach increases to $z \sim 1.5$ for
observed total
masses of $\sim 400\,M_\odot$. 
The ET will be able to explore even more remote distances, reaching
to $z\sim 5$ and further. Our present 
knowledge of stellar formation at such large redshifts is incomplete,
therefore we have computed the event rates for the ET integrating only
until $z=5$. We have compared three star formation models and three
different configurations of the binary. The effect of the particular
formation model 
is not very significant; remarkably, the physical
configuration does however influence the final rates. We
provide the results for 
three physical configurations, taking into account both single-
and double-cluster channels in the binary formation. For a fully correct
calculation of the event rate integrated over all possible
configurations, more detailed 
knowledge of the distribution of spins and mass ratios of IMBH
binaries formed in globular cluster is required. 

Advanced ground-based detectors are designed to be able to operate in different
modes so that their sensitivity can be tuned to various kinds of astrophysical
objects. Considering the importance of an eventual detection of a BBH,
the design of an optimized Advanced LIGO configuration for systems with $M \sim
10^{2-4}\,M_{\odot}$ would be desirable to increase the possibility of
observing such a system.  In case an IMBH binary coalescence were detected, the
recovery and study of the physical parameters of the system could serve to test
general relativity and prove or reject alternative theories, such as
scalar-tensor type or massive graviton theories. 
The {\em direct} identification of an IMBH with GWs will be a revolutionary
event not only due to the uncertainty that surrounds their existence and their
potential role in testing general relativity. The information encoded in the
detection will provide us with a detailed description of the environment of the
BBH/IMBH itself.

\acknowledgments 

We are indebted to C.~Cutler, B.~Sathyaprakash and B.~Krishnan for
discussions about the work. 
We thank C.~Reisswig and D.~Pollney for providing us with their
Cauchy-extrapolated data and F.~Ohme for helping us to implement the
post-Newtonian formalism used for the waveform modeling. We are grateful to
E.~Robinson and D.~Vanecek for comments on the manuscript.  PAS work
was partially supported 
by the DLR (Deutsches Zentrum f\"ur Luft- und Raumfahrt).  LS has been
partially supported by DAAD grant A/06/12630 and thanks the IMPRS for
Gravitational Wave Astronomy for support.

\end{document}